# A Novel Effective, Secure and Robust CDMA Digital Image Watermarking in YUV Color Space Using DWT2

Mehdi Khalili[1]

[1] Institute for Informatics and Automation Problems, National Academy of Science
Yerevan, Armenia
*Khalili@ipia.sci.am*

**Abstract**
This paper is allocated to CDMA digital images watermarking for ownership verification and image authentication applications, which for more security, watermark $W$ is converted to a sequence and then a random binary sequence $R$ of size $n$ is adopted to encrypt the watermark; where $n$ is the size of the watermark. This adopting process uses a pseudo-random number generator to determine the pixel to be used on a given key. After converting the host image to YUV color space and then wavelet decomposition of Y channel, this adopted watermark is embedded into the selected subbands coefficients of Y channel using the correlation properties of additive pseudo- random noise patterns. The experimental results show that the proposed approach provides extra imperceptibility, security and robustness against JPEG compression and different noises attacks compared to the similar proposed methods. Moreover, the proposed approach has no need of the original image to extract watermarks.

## 1. Introduction

Nowadays, watermarking of images is becoming increasingly of interest in tasks such as copyright control, image identification, verification, and data hiding [4]. Advances in computer networking and high speed computer processors have made duplication and distribution of multimedia data easy and virtually costless, and have also made copyright protection of digital media an ever urgent challenge. As an effective way for copyright protection, digital watermarking, a process which embeds (hides) a watermark signal in the host signal to be protected, has attracted more and more research attention [1-3]. A variety of watermarking techniques has been proposed in recent years. One of the earlier watermarking techniques, which used wavelet transform, was based on the adding pseudo random codes to the large coefficients at the high and middle frequency bands of the discrete wavelet transform [6]. This paper is allocated to CDMA digital images watermarking for ownership verification and image authentication applications, which for more security, watermark $W$ is converted to a sequence and then a random binary sequence $R$ of size $n$ is adopted to encrypt the watermark, where $n$ is the size of the watermark. This adopting process uses a pseudo-random number generator to determine the pixel to be used on a given key. This adopted watermark embeds into the selected subbands coefficients of Y channel of YUV color space of host image using the correlation properties of additive pseudo-random noise patterns. Obtained results of the experiments show the efficiency of proposed technique in acceptable transparency, high security and robustness against jpeg compression and different noise attacks in comparing earlier works such as [5].

## 2. YUV Color Space

The YUV color space is widely used in video and broadcasting today. It is very different from RGB color space; instead of three large color channels, it deals with one brightness or luminance channel (Y) and two color or chrominance channels (U-blue and V-red). The transformation from RGB to YUV that retains the same number of colors in both spaces is [9]:

$$\begin{bmatrix} Y \\ U \\ V \end{bmatrix} = \begin{bmatrix} 0.299 & 0.587 & 0.114 \\ -0.147 & -0.289 & 0.436 \\ 0.615 & -0.515 & -0.100 \end{bmatrix} \begin{bmatrix} R \\ G \\ B \end{bmatrix} \quad (1)$$

while the inverse conversion can be achieved with [9]:

$$\begin{bmatrix} R \\ G \\ B \end{bmatrix} = \begin{bmatrix} 1.000 & 0 & 1.140 \\ 1.000 & -0.395 & -0.581 \\ 1.000 & 2.032 & 0 \end{bmatrix} \begin{bmatrix} Y \\ U \\ V \end{bmatrix} \quad (2)$$

## 3. CDMA

CDMA (Code Division Multiple Access) is a form of spread spectrum where the signal i.e., watermark information, is transmitted on a bandwidth much greater than the frequency content of the original information, in this case, an image. In other words the transmitted signal bandwidth is much greater than the information bandwidth. Spread spectrum uses wide-band, noise-like signals, hence making it hard to detect. The band spread is accomplished by a pseudo-random code, which is independent of the data [14].

## 4. Proposed Watermarking Scheme

The current study task of digital watermarking is to make watermarks invisible to human eyes as well as robust to various attacks. The proposed watermarking scheme can hide visually recognizable patterns in images. This scheme is based on the discrete wavelet transform (DWT). Moreover, it has no need of the original host image to extract the embedded watermarks. RGB color space is highly correlated and is not suitable for watermarking applications [8]. So, in the proposed scheme, the host image is converted into YUV color space; the Y channel is then decomposed into wavelet coefficients. For more security of watermark, the watermark $W$ is converted to a sequence and then a random binary sequence $R$ of size $n$ is adopted to encrypt the watermark, where $n$ is the size of the watermark. This adopting process uses a pseudo-random number generator to determine the pixel to be used on a given key. The selected details subbands coefficients for embedding i.e. HL and LH coefficients are quantized and then their most significant coefficients are embedded by the adopted watermark using the correlation properties of additive pseudo-random noise patterns according to equation shown in below:

$$I_{Wx,y}(u,v) = \begin{cases} I_{x,y} + k * W_i & \text{if } W = 0 \\ I_{x,y} * W_i & \text{Otherwise} \end{cases} \quad (3)$$

In this equation $k$ denotes a gain factor for completely controlling the imperceptibility of watermarked images and the robustness of watermarks and also $I_W$ is the resulting watermarked image. This adaptive casting technique is utilized to embed the watermark coefficients for completely controlling the imperceptibility of watermarked images and the robustness of watermarks. To retrieve the watermark, after converting watermarked image from RGB color space to YUV color space, the Y channel will be decomposed into the wavelet coefficients. Then the same pseudo-random noise generator algorithm is seeded with the same key, and the correlation between the noise pattern and possible watermarked image in details subbands embedded coefficients will be computed. By computation of the each coefficient correlation whit a certain threshold T, the watermark is detected, and a single bit is set. The block diagram of the proposed watermarking scheme is shown in Figure (1).

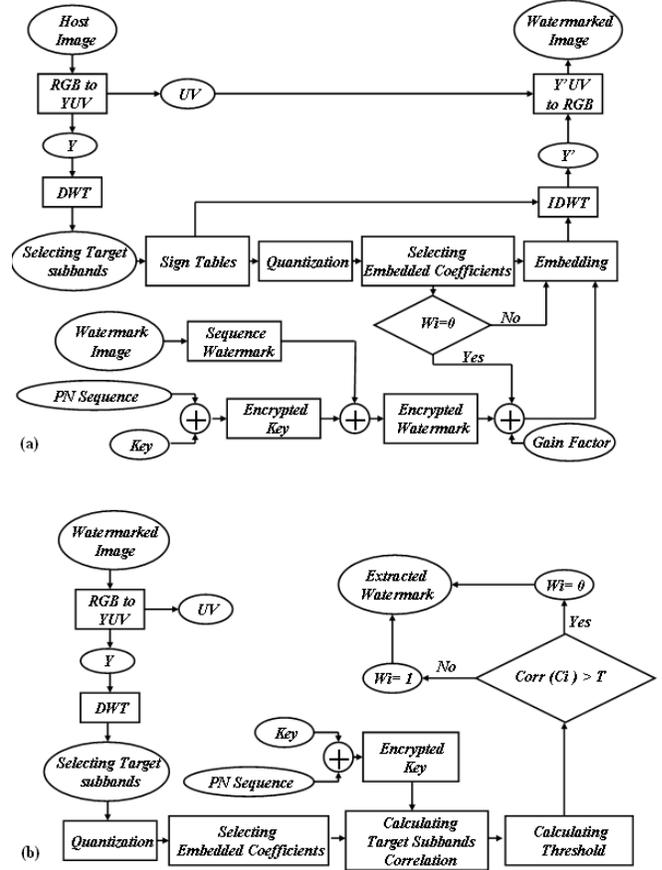

Fig. 1 Block diagrams of the proposed watermarking scheme; (a)Embedding procedure and (b)Extracting procedure.

### 4.1 Watermark Embedding Framework

The algorithm for embedding watermark in details subbands coefficients of the Y channel of host image is described as follows:

Step 1: Convert *RGB* channels of the host image *I* into *YUV* channels using Eq. (1).

Step 2: For more security of watermark, the watermark *W* is converted to a sequence and then a random binary sequence *R* of size *n* is adopted to encrypt the watermark, where *n* is the size of the watermark image. Then, the encrypted watermark sequence *W*1 is generated by a pseudo- random number generator to determine the pixel to be used on a given key.

Step 3: Decompose the *Y* channel into a one-level wavelet structure with four *DWT* subbands, *F(H)*. The coarsest coefficient of subbands *HL* and *LH* are taken as the target for embedding the watermark.

Step 4: Take absolute values on coefficients of all *LH* and *HL*, and record their signs in sign matrices.
Step 5: Quantize absolute values of selection coefficients.
Step 6: Embed encrypted watermark *W*1 into the coarsest coefficient of subbands *HL* and *LH* by the watermark embedding strategy shown in Eq. (3).
Step 7: Effect sign matrices into the embedded coefficients.
Step 8: Reconvert *YUV* channels of the changed host image into *RGB* channels.
Step 9: A watermarked image *I'* is then generated by inverse *DWT* with all changed and unchanged *DWT* coefficients.
Step 10: Record the pseudo-random noise generator algorithm and the key.

4.2 Watermark Extracting Framework

The embedded watermark in details subbands coefficients of the *Y* channel of host image is extracted using the same pseudo-random noise generator algorithm is seeded with the same key, and computation of the correlation between the noise pattern and possible watermarked image as follows:
Step 1: The *RGB* channels of the watermarked image are converted into *YUV* channels.
Step 2: Decompose the *Y* channel into four *DWT* subbands.
Step 3: Seeding the recorded key using the recorded pseudo-random noise generator algorithm.
Step 4: Quantize absolute values of *HL* and *LH* subbands.
Step 5: Computation of threshold *T* as follows:

$$T = \frac{Correlation(HL) + Correlation(LH)}{2} \quad (4)$$

Step 6: Computation of the threshold *T* and each embedded coefficient correlation, separately.
Step 7: The sequence watermark is extracted as follows:

$$\begin{cases} W_i = 0 & \text{if } C_i \rangle T \\ W_i = 1 & \text{Othetwise} \end{cases} \quad (5)$$

Step 8: The image watermark is produced by reconverting the extracted sequence watermark.

## 5. Experimental Results

Robustness is the most highly desired feature of a watermarking algorithm especially if the application demands copyright protection, and persistent owner identification.
The proposed perceptual watermarking scheme was implemented for evaluating both properties of imperceptibility and robustness against different attacks such as jpeg compression and noise.

Three 512×512 famous images: *Lena*, *Peppers* and *Baboon*, shown in Fig. 2(a-c) were taken as the host images to embed a 15×64 binary watermark image, shown in Fig. 2(d). For gain factor *k*, different values 0.5, 1.0 and 1.5 were taken entire implementation of the proposed CDMA scheme. For the entire test results in this paper, MATLAB R2007a software is used, too. 9 digit "key" used as initial state of MATLAB random number generator. Also for computation of the wavelet transforms, 9-7 biorthogonal spline (B-spline) wavelet filters are used. Cause of use of B-spline function wavelet is that, B-spline functions, do not have compact support, but are orthogonal and have better smoothness properties than other wavelet functions [6, 10].

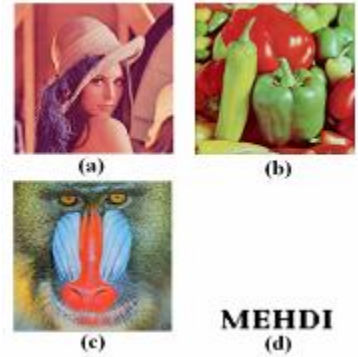

Fig. 2 Proposed (a-c) The host images (Lena, Peppers, and Baboon) and (d) The watermark image.

After watermark embedding process, the similarity of original host image *x* and watermarked images *x'* was measured by the standard correlation coefficient (Corr) as fallows [4, 7, 11]:

$$Correlation = \frac{\sum (x - x')(y - y')}{\sqrt{(x - x')^2}\sqrt{(y - x')^2}} \quad (6)$$

where *y* and *y'* are the discrete wavelet transforms of *x* and *x'*, respectively. Moreover, the peak signal-to-noise ratio (*PSNR*) was used to evaluate the quality of the watermarked image. The *PSNR* is defined as [4, 7, 11]:

$$PSNR = 10 \, log_{10} \frac{255^2}{MSE} \, (dB) \quad (7)$$

where mean-square error (*MSE*) is defined as [5, 7, 11 and 12]:

$$MSE = \frac{1}{mn} \sum_{i=1}^{m} \sum_{j=1}^{n} (h_{i,j} - h'_{i,j})^2 \quad (8)$$

Where $h_{i,j}$ and $h'_{i,j}$ are the gray levels of pixels in the host and watermarked images, respectively. The larger *PSNR* is, the better the image quality is. In general, a watermarked image is acceptable by human perception if its *PSNR* is greater than 30 dBs. In other words, the correlation is used for evaluating the robustness of watermarking technique and the *PSNR* is used for evaluating the transparency of watermarking technique [4, 7 and 11].

Also the normalized correlation (*NC*) coefficient was used to measure the similarity between original watermarks *W* and the extracted watermarks *W'* that is defined as [4, 7 and 11]:

$$NC = \frac{\sum_i \sum_j w_{i,j} * w'_{i,j}}{\sum_i \sum_j w_{i,j}^2} \qquad (9)$$

The proposed CDMA watermarking scheme yields satisfactory results in watermark imperceptibility and robustness. The embedding of large watermarks using CDMA requires the embedding gain *k* to be lowered to preserve the visual quality of the image. The obtained results show that larger gains are reason that CDMA will be remained as more *PN* sequences are added to the host image but it will be caused to decrease the transparency of the image, because of decreasing correlation between original image and watermarked image. Also, this results show that, the best compression can be made with CDMA, although CDMA is more resistant to different noise attacks such as Gaussian and salt & pepper. The PSNRs of the watermarked images produced by the proposed scheme for different gain factors *k* are all greater than 78 dBs and Normalized correlations (NC) between original watermark images and extracted watermark images are all equal 1. The results for effecting *k* on watermark imperceptibility are illustrated in figure (3) and table (1). The PSNR and correlation plots shown in figure (3) indicates that the proposed scheme satisfy imperceptibility as well. After able to achieve the desired fidelity, various attacks were performed to test the robustness of the proposed scheme and it was found that the proposed scheme performs excellently against JPEG compression and different noise attacks.

### 5.1 Robustness to JPEG compression

To evaluate the response of the watermarking scheme to JPEG compression, watermarked images were compressed with different JPEG qualities *Q*s: 10, 15, 25, 50 and 75. Figure (4) shows the JPEG compression of the watermarked images under quality factor 10 and gain factor 1. Figure (5) shows the extracted watermarks from watermarked images after JPEG compression under different qualities and different gain factors. To the similar references [5 and 12], the experimental results respectively under different JPEG qualities and gain factors are compared.

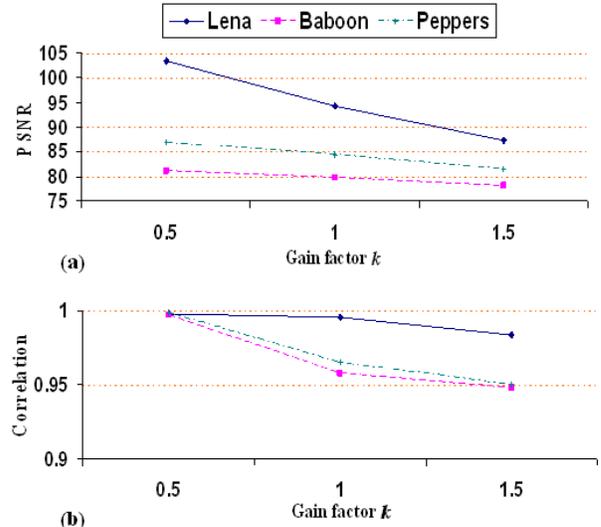

Fig. 3 (a) Influence of the gain factor *k* on correlation and (b) Influence of the gain factor *k* on PSNR.

Table 1: Obtained results of watermark extracting with different values of the gain factor *k*

| *Image* | *k* | *PSNR (dB)* | *Corr* | *NC* | *Error Bit %* |
|---|---|---|---|---|---|
| Lena | 0.5 | 106.07 | 0.9980 | 1.00 | 0 |
| | 1.0 | 94.30 | 0.9958 | 1.00 | 0 |
| | 1.5 | 87.42 | 0.9839 | 1.00 | 0 |
| Peppers | 0.5 | 86.81 | 0.9987 | 1.00 | 0 |
| | 1.0 | 84.36 | 0.9648 | 1.00 | 0 |
| | 1.5 | 81.48 | 0.9503 | 1.00 | 0 |
| Baboon | 0.5 | 81.13 | 0.9971 | 1.00 | 0 |
| | 1.0 | 79.77 | 0.9576 | 1.00 | 0 |
| | 1.5 | 78.03 | 0.9477 | 1.00 | 0 |

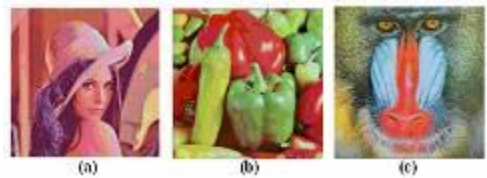

Fig. 4 Watermarked images under JPEG compression (Q=10, k=1); (a) Lena, (b) peppers and (c) Baboon.

| Image | K | Q | | | | |
|---|---|---|---|---|---|---|
| | | 10 | 15 | 25 | 50 | 75 |
| Lena | 0.5 | | | | MEHDI | MEHDI |
| | 1.0 | | | MEHDI | MEHDI | MEHDI |
| | 1.5 | | MEHDI | MEHDI | MEHDI | MEHDI |
| Peppers | 0.5 | | | | MEHDI | MEHDI |
| | 1.0 | | | MEHDI | MEHDI | MEHDI |
| | 1.5 | | MEHDI | MEHDI | MEHDI | MEHDI |
| Baboon | 0.5 | | | | MEHDI | MEHDI |
| | 1.0 | | MEHDI | MEHDI | MEHDI | MEHDI |
| | 1.5 | MEHDI | MEHDI | MEHDI | MEHDI | MEHDI |

Fig. 5 Extracted watermarks from watermarked images after JPEG compression with different values for Q and *k*.

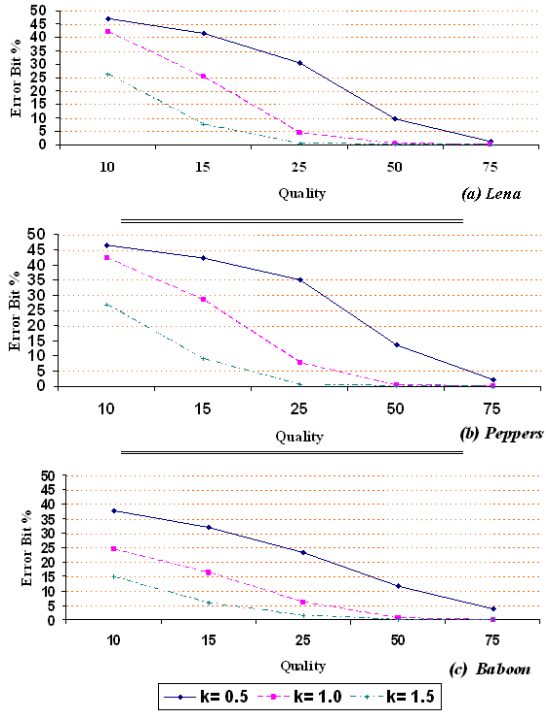

Fig. 6 The percentage of error bit in extracted watermarks in JPEG compression under deferent qualities and gain factors.

Table 2: Obtained results of extracted watermarks in JPEG compression

| k=0.5 | | | | | | |
|---|---|---|---|---|---|---|
| Image | Q | 10 | 15 | 25 | 50 | 75 |
| Lena | NC | 0.0575 | 0.1753 | 0.4066 | 0.8107 | 0.9725 |
| | PSNR | 91.07 | 93.48 | 96.28 | 99.24 | 97.25 |
| Peppers | NC | 0.067 | 0.1591 | 0.3123 | 0.7367 | 0.9527 |
| | PSNR | 87.35 | 89.40 | 91.48 | 93.83 | 95.91 |
| Baboon | NC | 0.2608 | 0.3728 | 0.5629 | 0.7681 | 0.9210 |
| | PSNR | 81.05 | 82.95 | 85.25 | 88.65 | 92.54 |
| k=1.0 | | | | | | |
| Image | Q | 10 | 15 | 25 | 50 | 75 |
| Lena | NC | 0.2107 | 0.5114 | 0.9094 | 0.9953 | 1.00 |
| | PSNR | 87.49 | 89.14 | 91.05 | 93.65 | 97.98 |
| Peppers | NC | 0.1591 | 0.4538 | 0.8475 | 0.9953 | 1.00 |
| | PSNR | 84.85 | 86.39 | 88.03 | 90.31 | 93.41 |
| Baboon | NC | 0.5278 | 0.6828 | 0.8736 | 0.9838 | 1.00 |
| | PSNR | 79.87 | 81.58 | 83.73 | 87.14 | 91.56 |
| k=1.5 | | | | | | |
| Image | Q | 10 | 15 | 25 | 50 | 75 |
| Lena | NC | 0.4996 | 0.8459 | 0.9907 | 1.00 | 1.00 |
| | PSNR | 83.74 | 84.98 | 86.65 | 89.85 | 95.66 |
| Peppers | NC | 0.4761 | 0.8214 | 0.9884 | 0.9977 | 1.00 |
| | PSNR | 81.87 | 83.05 | 84.58 | 87.37 | 91.90 |
| Baboon | NC | 0.7147 | 0.8825 | 0.9681 | 1.00 | 1.00 |
| | PSNR | 78.24 | 79.81 | 81.86 | 85.63 | 90.87 |

The percentage of the error bit for extracted watermarks under JPEG compression is shown in figure (6). As it is seen, the percentage of error bit plot shown in Fig. (6) indicates that the margin of error is very less for the detection statistic, and CDMA proposed scheme has increased in comparing with the earlier works such as [5 and 12]. Table (2) shows the response of the detector to extract the watermark in JPEG compression.

From the obtained results it can be seen that for JPEG compression with a quality factor of 75 and gain factor 0.5, a quality factor of 50 and gain factor 1 and also a quality factor of 15 and gain factor 1.5, the watermark detection and extraction is near perfect. The recovered watermarks for a quality factor of 25 and gain factor 1 show a number of detection errors and these only become highly noticeable for a quality factor of 15 and gain factor 1. The watermarks are still recognizable for a quality factor of 10 and gain factor 1. The overall robustness of proposed scheme for JPEG compression is considered high level, according to the robustness requirements table provided by Petitcolas [13] and is higher than earlier works such as [5]; therefore, It can be said that the proposed scheme is more robust than earlier works such as [5] against JPEG compression, even for low JPEG quality.

### 5.2 Robustness to Noise Attacks

The CDMA proposed scheme was tested for its robustness against different types of noise. This is done by first introducing noise into the watermarked images. Gaussian noise with zero mean is introduced to verify as to what extent the proposed scheme can withstand noise. From the results shown below, it is observed that for a Gaussian noise of 1% with the gain factor 0.5; the watermark recovery is almost recognizable, for a Gaussian noise of 1% with the gain factor 1, the watermark recovery is moderate and for a Gaussian noise of 1% with the gain factor 1.5, the watermark recovery is near perfect with very few detection errors. We must keep in mind that most schemes offer moderate robustness to noise [14].

Figure (7) shows added Gaussian noise of 0.5% with gain factor 1 to watermarked images. Figure (8) shows the extracted watermarks in Gaussian noise of 1% with different gain factors.

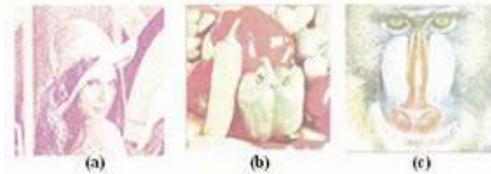

Fig. 7 Watermarked images under Gaussian noise of 0.5% with gain factor 1; (a) Lena, (b) Peppers and (c) Baboon.

Figure (9) shows the percentage of error bit in extracted watermarks under different variances and different gain factors. Figure (10) shows the results of PSNR in Gaussian noise experiment. As it is obvious, the lowest value for PSNR is still greater than 49 db (Gaussian noise of 1% with gain factor 0.5). Figure (11) shows the results of NC in Gaussian noise experiment. It is visible, the NC is acceptable for Gaussian noise of 1% with the gain factor 1.5 and it is moderate in Gaussian noise of 1% with the gain factor 1. The obtained results from Gaussian noise experiment show that, the proposed CDMA scheme is more robust than the earlier works such as [5].

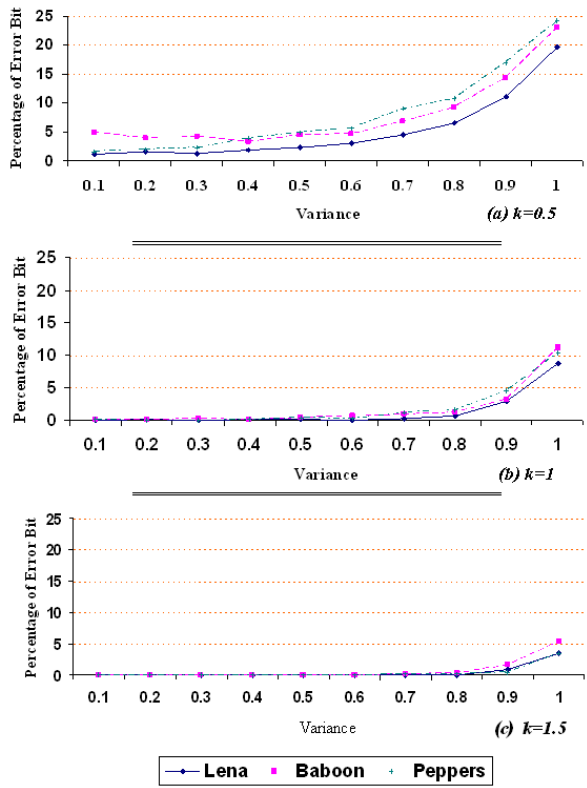

Fig. 8 Extracted watermarks from Gaussian noise of 1% with different gain factors; (a) Lena, (b) Peppers and (c) Baboon.

Fig. 9 Percentage of error bit under Gaussian noise experiment.

When the salt & pepper noise with zero mean and different noise densities 0.01 to 0.5 with different gain factor introduced in the watermarked images, the extracted watermarks are recognizable in gain factor 0.5 and noise density 0.5, they are recovered in gain factor 1 and noise density 0.5 moderately and they are recovered in gain factor 1.5 and noise density 0.5 with a few detection errors.

Figure (12) shows the watermarked images under salt & pepper noise attacks with noise density 0.5 and gain factor 1. Figure (13) shows the extracted watermarks from noisy watermarked images under noise density 0.5 with different gain factors and figure (14) shows the percentage of error bit in extracted watermarks under salt & pepper noise with different noise densities and different gain factors. The PSNR results in salt & pepper noise experiment is shown in figures (15). As it is obvious, the lowest value for PSNR is still greater than 50 (noise density 0.5 with gain factor 0.5). Figure (16) shows the NC results in salt & pepper noise experiment. The obtained results show that, NC in salt & pepper noise experiment is acceptable for noise density 0.5 with the gain factor 0.5, it is moderate for the same noise density with the gain factor 1 and it is near perfect for the same noise density and gain factor 1.5 with a few detection errors. From the obtained results it can be said that the proposed CDMA scheme is very efficient in robustness against salt & peppers attacks and it improves the results in the earlier works such as [5].

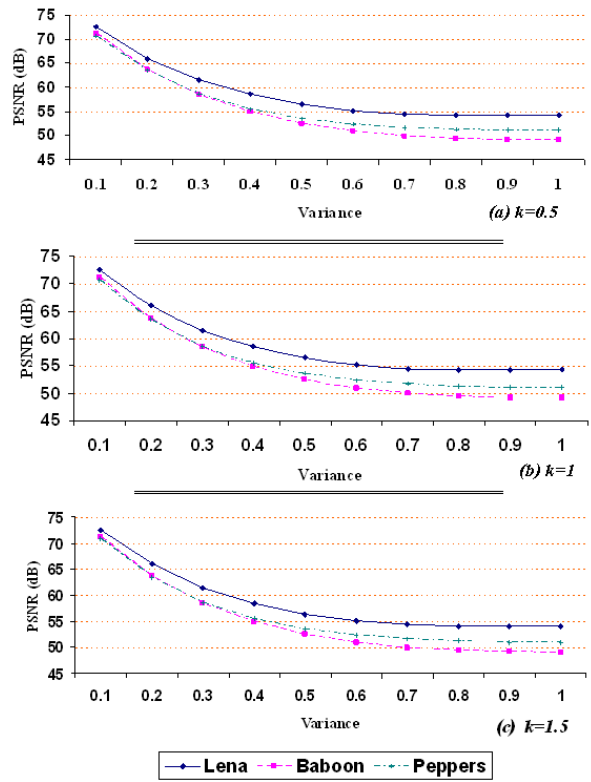

Fig. 10 PSNR under Gaussian noise experiment.

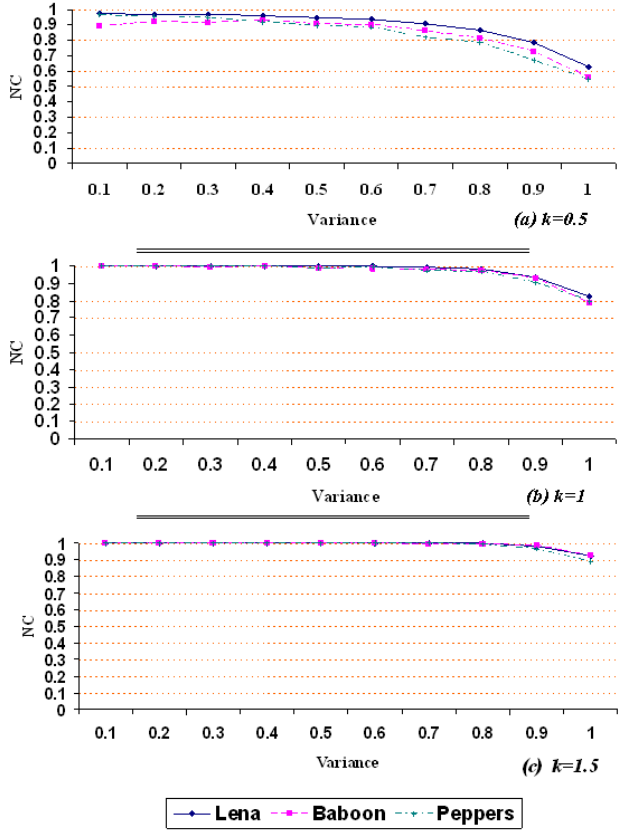

Fig. 11  NC in Gaussian noise experiment.

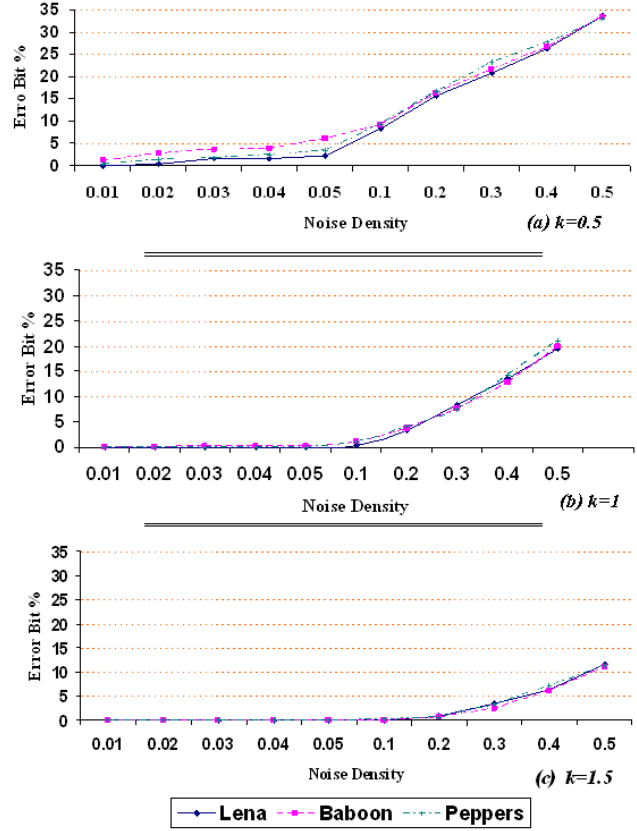

Fig. 14  Percentage of error bit under salt & pepper noise experiment.

Fig. 12  Salt & pepper noise on watermarked images with variance 0.5 and gain factor 1;  (a) Lena, (b) Peppers and (c) Baboon.

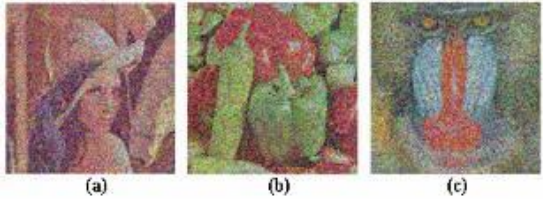

Fig. 13  Extracted watermarks from salt & pepper noise with noise density 0.5different gain factors.

## 6. Conclusions

This paper provides a novel CDMA watermarking scheme using DWT2 with satisfactory results in watermark imperceptibility, robustness and security that improves upon the earlier work such as [5].

In the scheme, the host image is converted into *YUV* channels; then, the *Y* channel is decomposed into wavelet coefficients. For more security of watermark, the watermark *W* is converted to a sequence and then a random binary sequence *R* of size *n* is adopted to encrypt the watermark, where *n* is the size of the watermark using a pseudo-random number generator to determine the pixel to be used on a given key. The selected details subbands coefficients for embedding are quantized and then their most significant coefficients are replaced by the adopted watermark using the correlation properties of additive pseudo-random noise patterns. To embed the watermark coefficients for completely controlling the imperceptibility and the robustness of watermarks, an adaptive casting technique is utilized using a gain factor *k*. Also, the CDMA watermark scheme has no need to original Image in extracting process.

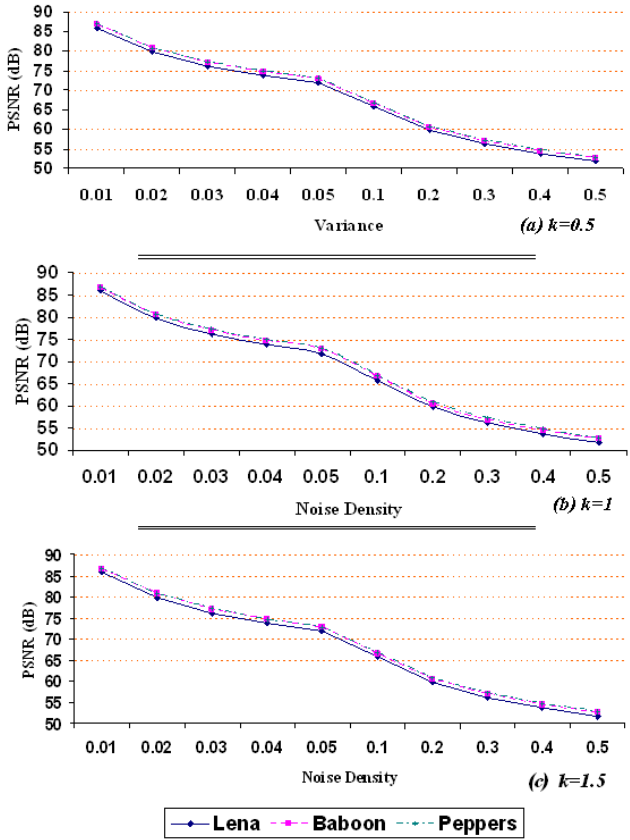

Fig. 15 PSNR under salt & pepper noise experiment.

The observations regarding the proposed watermarking scheme are summarized as follows: (1) Increasing gain factor $k$ increases the PSNR and NC. In a result, it decreases the percentage of error bit and increases the robustness of the watermark against JPEG compression and different noise attacks such as Gaussian and salt & pepper. In opposite, increasing gain factor $k$, decreases the transparency property.(2) The robustness of the proposed scheme to JPEG compression is found to be very good at a quality factor of 75 and gain factor 0.5, a quality factor of 50 and gain factor 1 and also a quality factor of 15 and gain factor 1.5. Reasonably, the robustness of the proposed scheme to JPEG compression is found to be good at a quality factor of 25 and gain factor 1 with a number of detection errors and for a quality factor of 15 and gain factor 1, this becomes highly noticeable. (3) The results show that, the watermark is still recognizable for a quality factor of 10 and gain factor 1. These results show that, the proposed scheme improves the results in earlier works such as [5]. (4) The robustness of the proposed scheme to Gaussian noise with zero mean is found to be very good for a Gaussian noise of 1 % with the gain factor 1.5, and it is good for a Gaussian noise of 1 % with the gain factor 1, and it is recognizable for a Gaussian noise of 1 % with the gain factor 0.5. This results show the improving in comparing with the earlier works such as [5]. (5) The robustness of the proposed scheme to salt & pepper noise with zero mean with noise density 0.5 is found to be very good for a gain factor 1.5, it is good for the same noise density and a gain factor 1 and it is recognizable for the same noise density and a gain factor 0.5.

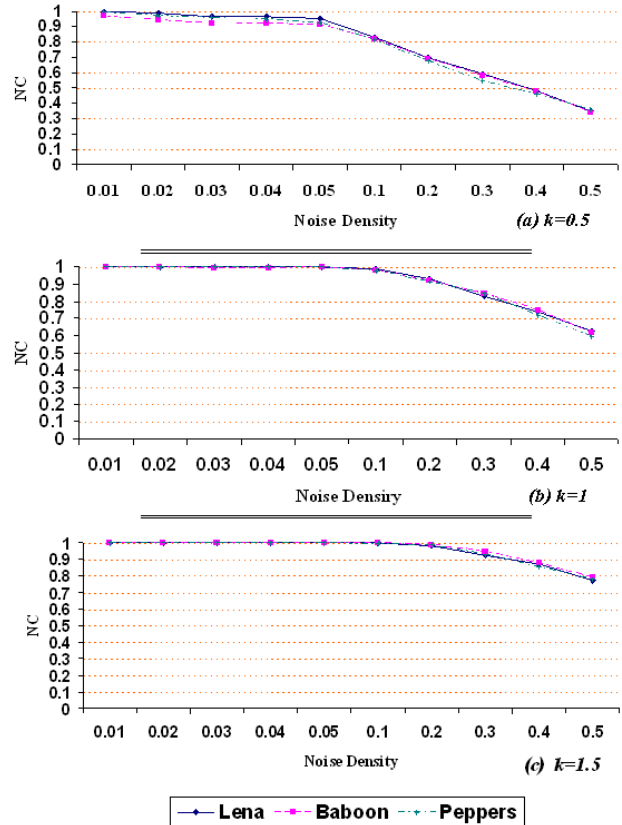

Fig. 16 NC in salt & pepper noise experiment.

## References

[1] Jian. Ren, "A cryptographic watermarking technique for multimedia signals", Springer Science, Business Media, August 2008.
[2] Wang, S.-H. and Lin, Y.-P, "Wavelet tree quantization for copyright protection watermarking", IEEE Trans, Image Process 13(2), pp.154–165, February 2004.
[3] Malvar, H.S. and Florêncio, D.A.F, "Improved spread spectrum: a new modulation technique for robust watermarking", IEEE Trans. Signal Proc. 51(4), pp.898–905, April 2003.
[4] Mehdi Khalili and David Asatryan, "Effective Digital Image Watermarking in YCbCr Color Space Accompanied by Presenting a Novel Technique Using DWT", Mathematical Problems of Computer Science 33, pp.150-161, 2010.
[5] Yanmei Fang, Jiwu Huang and Yun Q. Shi, "Image Watermarking Algorithm Applying CDMA", IEEE, in Proc. ISCAS (2), pp.948-951, 2003.

[6] X.G. Xia, C. G. Boncelet and G. R. Arce, "Wavelet transform based watermark for digital images", Optics Express, Vol. m No. 12, pp.497-511, Dec. 1998.

[7] M. Khalili, "A Comparison between Digital Images Watermarking in Two Different Color Spaces Using DWT2", CSIT, Proceedings of the 7th International Conference on Computer Science and Information Technologies, Yerevan, Armenia, pp. 158-162, 2009.

[8] Qian-Chuan Zhong, Qing-Xin and Ping- Li- Zhang, "A Satial Domain Color Watermarking Scheme Based on Chaos", IEEE, 2008.

[9] Annajirao Garimella, M. V. V. Satyanarayana, P.S. Murugesh and U.C. Niranjan, "Asic for Digital Color Image Watermarking", 11th Digital Signal Processing Workshop & IEEE Signal Processing Education Workshop, 2004.

[10] Chiunhsiun Lin, "Face detection in complicated backgrounds and different illumination conditions by using YCbCr color space and neural network", Elsevier, Pattern Recognition Letters 28 (2007), pp. 2190–2200, July 2007.

[11] Ming-Shing Hsieh, "Wavelet-based Image Watermarking and Compression", Ph.D Thesis, Institute of Computer Science and Information Engineering National Central University, Taiwan, Dec. 2001.

[12] William A. Irizarry-Cruz, "FPGA Implementation of a Video Watermarking Algorithm", M.S. Thesis, University of Puerto Rico Mayaguez Campus, 2006.

[13] F. A. P. Petitcolas, "Watermarking schemes evaluation." IEEE Signal Processing, vol. 17, no. 5, pp. 58–64, September 2000.

[14] Arvind Kumar Parthasarathy, "Improved Content Based Watermarking for Images", M.S. Thesis, Louisiana State University and Agricultural and Mechanical College, Kanchipuram, India, August 2006.